\documentclass[twocolumn,aps,prl,reprint]{revtex4}

\usepackage{graphicx}
\usepackage{float}
\graphicspath{./}

\usepackage{multirow}

\usepackage{hyperref}

\usepackage[T1]{fontenc}

\begin{document}

\title{Confinement of wave-function in Fractal geometry, a detection using DFT}
\author{Mohammed Ghadiyali}
\email{ghadiyali.mohd@physics.mu.ac.in}
\author{Sajeev Chacko}
\email{sajeev.chacko@physics.mu.ac.in; \\sajeev.chacko@gmail.com}
\affiliation{Department of Physics, University of Mumbai,
	\\Kalina Campus, Santacruz (E), Mumbai - 400 098, India.} \

\begin{abstract}
Fractals are self-repeating patterns which have dimensions given by fractions rather than
integers.
While the dimension of a system unambiguously defines its properties, a fractional dimensional system
can exhibit interesting properties.
The recent work on confinement of the electronic wavefunction in fractal dimensions by creating
artificial lattice has given rise to new possibilities of designing artificial lattice.
In this study, we demonstrate that the first principle methods can be effectively employed to investigate, design and
characterize electrons in Hausdorff dimension.
We apply the method to study the molecular graphene and Lieb lattices which lead to fractal lattices based on these fractals.
We construct the hexaflake and Vicsek fractals by adsorption of CO molecules on Cu(111) surface.
This opens up the possibility of using high throughput techniques for screening and discovering
such lattices from currently known crystals or developing them altogether.
\end{abstract}

\maketitle

Dimension of an object is defined by the number of coordinates required for specifying a point 
on it.
For example, a point residing on a line requires a unit coordinate, hence its dimension is one,
while a point on a cube would require three coordinates, so its dimension is three.
An alternate definition is provided by Hausdorff, called as ``Hausdorff Dimension'' where
the log log plot of two descriptors is taken as dimension~\cite{Hausdorff_dimension}.
This can be implemented via box counting or correlation or just on the basis of information 
needed to identify features based on probability or others.

Fractional dimensions are used for explaining self repeating mathematical patterns named as
``Fractals''.
However, these fractals have also emerged as important constructs which have found applications
such as stretchable electronics~\cite{Frac_app_1}, hydrogen storage~\cite{Frac_app_2},
transistors~\cite{Frac_app_3}, antennas~\cite{Frac_app_4, Frac_app_5, Frac_app_7}, medical
imaging~\cite{Frac_app_6}, etc.
Apart from these, fractals are observed as Hofstadter butterfly - a behaviour of electrons under
perpendicular magnetic fields~\cite{Hofstadter_butterfly_1},~\cite{Hofstadter_butterfly_2},~\cite{Hofstadter_butterfly_3} and also exhibits fractional quantum Hall effect~\cite{Fractal_QHE},
in splitting of energy levels explained via fractal dimensions~\cite{Fractal_energy_states}, in 
quantum transport~\cite{Qtrnas_Fractal_1, Qtrnas_Fractal_2}, etc.
Also, fractals are observed in self assembled polymers~\cite{self_ass_1, self_ass_2, self_ass_3,
self_ass_4}.

Recently, Smith~\textit{et al.} reported that they have confined the wavefunction of carbon
monoxide molecules~(CO) on Cu(111) surface patterned as Sierpiński triangle in fractal
dimensions~\cite{Fractal_Confinement}.
For performing this work, Smith~\textit{et al.} use a scanning tunnelling microscope~(STM) for
manipulation of CO on Cu(111).
This demonstration of confinement of wavefunctions in non-integer dimensions opens up 
possibilities of developing new applications.
To facilitate this, a detailed understanding of the same is required.
Here, we demonstrate that using \textit{ab initio} calculations within the framework 
of density functional theory~(DFT) can be applied for designing and characterization of 
electrons in fractal geometry.

Density functional theory is better suited methodology as it can be combined with high throughput 
techniques to screen large number of crystals based on their topology.
This could allow for detection of systems which inherently demonstrate crystal structures 
having fractal geometry.
Further, the comprehensiveness and competence of DFT not only allow it to investigate different schemes for artificial lattices based on fractals but also their applications.
This can be performed using other methodology also, however, the ease and efficiency of DFT are exceptionally strong.

As an example, the application pointed out by Smith~\textit{et al.} for using fractal lattices as a controlled environment while performing spectroscopy can be achieved {\em in silico} by DFT. 
In other recent work by Fabian~\textit{et al.}, an atom of Holmium was used to create the smallest magnet~\cite{Single_atom_magnet}.
This combined with fractal lattice can be used to create {\em Halbach-like} array in which the magnetic field on one side of the array is nearly zero~\cite{Halbach_magnet}.
Such systems can be studied extensively {\em in silico} by DFT.
    
The work presented is performed via \textsc{Quantum ESPRESSO} (\textsc{QE}) package~\cite{QE}.
It is a plane wave pseudopotential based package.
The project augmented wave~(PAW) pseudopotentials~\cite{PAW} in the form of Perdew-Burke-
Ernzerhof (PBEsol) exchange correlation were used as available from the SSSP 
library~\cite{SSSP_1, SSSP_2}.
For improve accuracy van der Waals correction were also taken into consideration via DFT-D2 
methodology as implemented in QE.
The kinetic energy cutoff was set at 50Ry with a Monkhorst-Pack \textit{k}-mesh, each 
\textit{k}-point of the mesh was at a distance of 0.20~\AA$^{-1}$.
While, for calculating density of states, a finer mesh of 0.15~\AA$^{-1}$, was used.
A vacuum of 10\AA~was added along the z-axis to remove the interactions with the periodic images.

We select hexaflakes~\cite{Frac_app_7} and Vicsek fractals~\cite{Vicsek} for this study.
A hexaflake is an iterative fractal constructed by dividing a hexagonal flake in seven segments 
which may be described as a defective molecular graphene~\cite{Mol_graphene}.
On the other hand, Vicsek fractal is constructed by decomposition of squares in a 3 by 3 grid.
This system is described as defective Lieb lattice~\cite{Leib_lattice}.
As CO molecules adsorbed on Cu(111) surface have been used to construct both molecular 
graphene~\cite{Mol_graphene} as well as Lieb lattice~\cite{Leib_lattice}, it was the obvious 
choice for this study (see supplementary information).
Other possible constructs are Fe adatom on Cu(111)~\cite{STM_1}, Cl vacancy on chlorinated 
Cu(100)~\cite{STM_5} etc~\cite{STM_2, STM_3, STM_4}.
For constructing the systems both Hexaflake and Vicsek fractal, CO molecules are placed on three atom thick Cu(111) surface.
This surface is cleaved from bulk using ATK VNL Builder~\cite{VNL-2017_1} package.
The bottom two atomic layers were fixed and the system was allowed to relax along the direction 
of vacuum.
This approach of creating system is a direct derivative of the approach used by Paavilainen {\it 
et al.}~\cite{STM_4}, here a Kagome lattice of CO molecule was created on Cu(111) surface (it is further explained in supplementary information).

\begin{figure}[H]
	\centering
	\includegraphics[scale=0.4]{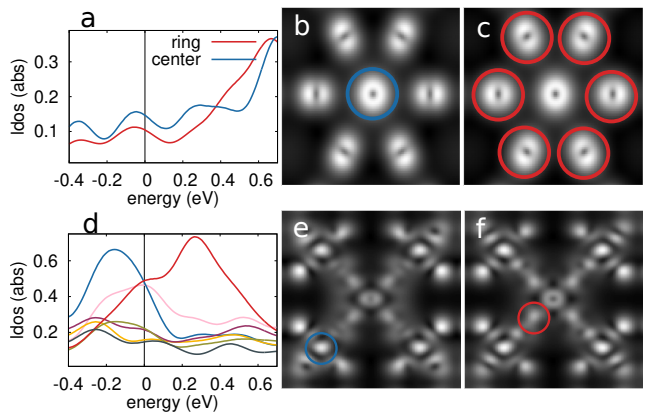}
	\caption{(colour online) Figures  (a) and (d) are the local density of states of systems
	    hexaflake and Vicsek fractal respectively.
	    The STM maps of hexaflake are given	in (b) and (c).
	    Their biases were kept at -0.3~eV and 0.5~eV respectively.
	    Similarly, (e) and (f) represent STM maps of Vicsek fractal at bias of -0.3~eV and 0.45~eV respectively. The colour in both the images represents the unique sites in both the fractals.
	}
	\label{DOS}
\end{figure}

In this work, we study the second generation of hexaflake and Vicsek fractals.
First generation of hexaflake is a single hexagon, while the second generation is
collection of seven, {\it i.e.} a graphene ring with a central atom as illustrated in 
figure~\ref{DOS}(a). 
Similarly, the second generation of Vicsek fractal is illustrated in figure~\ref{DOS}(d).
The theoretical Hausdorff dimensions of these fractals are 1.77 and 1.46 respectively.
The local density of states accompanying the STM maps are also given in figure~\ref{DOS}.
The STM maps are computed using the method proposed by Tersoff and Hamann~\cite{STM_theory} as implemented in QE package.

There are the six CO molecules forming the outer ring of hexaflake and central CO molecule.
From the LDOS of hexaflake (figure~\ref{DOS}(a)), it can be observed that this structure is
a composed of two types of CO molecules.
However, the density distribution between them is not very well resolved as they follow 
the same trend of increasing in density as the function of energy.
This is also demonstrated from the STM maps taken at two different biases of -0.3~eV and 
0.5~eV, where no change is observed.
This indicates that the hexaflake so formed is quite homogeneous.
On the other hand, a similar inference cannot be drawn for the Vicsek fractal.
From its LDOS, it may be clearly observed that the CO molecules are indifferent from each other,
when being grouped together.
However, except for two molecules, the density distribution for most of them is not well 
resolved.
This can be observed in the STM maps given in figure~\ref{DOS}(e) and~\ref{DOS}(f) (marked with 
blue and red circle).
The maps are computed for the bias of -0.3~eV and 0.45~eV respectively.
These same colors are also used to represent the LDOS plots in figures~\ref{DOS}(a) 
and~\ref{DOS}(d).

The dimensional analysis of both the systems was performed using box counting method.
The Hausdorff dimensions determined by this method are also known as Minkowski – Bouligand 
dimension.
This method is well suited as each pixel in the STM maps can be taken as a box of length 
$l_{box}$, and the number of boxes as $N(l_{box})$. Hence, the Hausdorff dimension would 
be~\cite{Hausdorff_dimension}:

$$D=\lim_{l_{box}\to0}\frac{\log (N(l_{box}))}{\log(l_{box})}$$

\begin{figure}[H]
	\centering
	\includegraphics[scale=0.4]{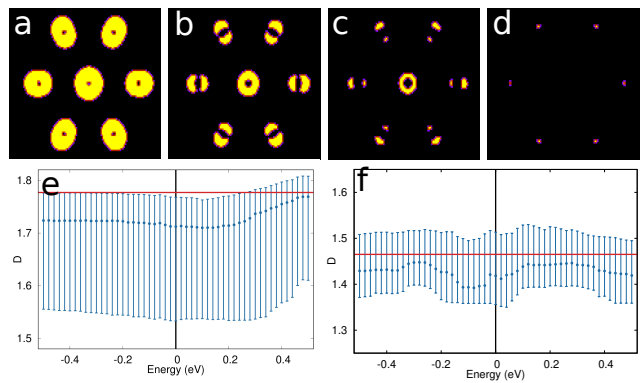}
	\caption{(colour online) In the figure above (a), (b), (c) and (d) are the STM maps of
		hexaflake taken at threshold percentage of 20\%, 50\%, 70\% and 90\% respectively, 
		while the plots (e) and (f) show the dimensional analysis done by box counting method 
		for hexaflake and Vicsek fractals respectively.
		The error bars represent the upper and lower bound of the fractal dimension.
		The red line represents actual fractal dimension for both systems at 1.77 and 1.46 for 
		hexaflake and Vicsek fractals respectively.
	}
	\label{FD}
\end{figure}

However, as the STM maps are not flat and have gradients, it is difficult to calculate the 
Hausdorff dimension at the boundaries.
To solve this problem we remove the gradients by setting binary values of the STM 
maps: for values above a certain threshold the STM map was set to 1, and 0 elsewhere.
In figure~\ref{FD}(a-d), the STM maps for hexaflake are given for threshold percentages of 
20\%, 50\%, 70\% and 90\% respectively.

The STM maps were calculated for the bias from -0.5 to 0.5~eV, the range in which the 
wavefunction is generally well-confined for CO on Cu(111).
In figure~\ref{FD}(e) and~\ref{FD}(f), the plot of Hausdorff dimension~(D) v/s bias is given.
The red horizontal line represents the Hausdorff dimensions for hexaflake and Vicsek fractal at
values 1.77 and 1.46, respectively.
The error bars in figure~\ref{FD} represent the upper and lower bounds of the calculated 
Hausdorff dimension at different threshold percentages.
For hexaflake, the threshold percentage is 20\% for upper bound, 70\% for lower bound and 30\% 
for the central range.
For Vicsek fractal, the threshold percentage for the central region is 80\% while the upper 
and lower bounds are given by 65\% and 90\% respectively.

Although the Hausdorff dimensions calculated differ from the know values of the structure, it 
can be observed that they are still non-integer values.
This difference is partially due to the approximation initially taken while computing 
the wavefunctions using DFT.
The other reason is the approximate construction of fractal geometry via CO on Cu(111) - 
calculations when performed on higher orders of the fractals may improve the results.
Nevertheless, it can be observed that both of the artificial lattices are in Hausdorff 
dimensions.

However, as this method of counting dimension is based just on the image processing techniques,
it can not alone be used with certainty to determine that the electron wavefunction is indeed in Hausdorff dimensions.
To demonstrated this point we adsorb benzene molecule on Cu(111) surface.
Upon the calculation of fractal dimension, according to the methodology describe above, it is found to be in between 1.85 to 1.76 (see supplementary information).
Hence, this type of analysis is not sufficient.

To determine if there the electron wavefunctions are in fractal dimensions, Lowdin charge analysis is performed.
The procedure for doing so is, first Lowdin charges of the complete system is calculated, then a second Lowdin charge calculation is performed with the substrate removed.
If no change is observed within the accuracy of DFT, then it can inferred that electron wavefunctions are in Hausdorff dimensions.
From the data presented in the table~\ref{Tab_lowdin_charges}, a change is observed for the Benzene on Cu(111), this is an indication of charge transfer between them, while for both the fractals this kind of change is not observed.

\begin{table}[]
	\begin{tabular}{|c|c|c|c|}
		\hline
		\multirow{2}{*}{\begin{tabular}[c]{@{}c@{}}Name of \\ \\ the System\end{tabular}} & \multicolumn{3}{c|}{Lowdin Charges}                                                                                                                                                         \\ \cline{2-4} 
		& \begin{tabular}[c]{@{}c@{}}Complete\\ System\end{tabular} & \begin{tabular}[c]{@{}c@{}}Isolated\\ System\end{tabular} & \begin{tabular}[c]{@{}c@{}}Percentage \\ Change\\ (\%)\end{tabular} \\ \hline
		Benzene on Cu                                                                     & 29.25                                                     & 29.54                                                     & 0.991                                                               \\ \hline
		Hexaflake                                                                         & 69.05                                                     & 69.15                                                     & 0.144                                                               \\ \hline
		Vicsek fractal                                                                    & 246.19                                                    & 245.95                                                    & 0.0975                                                              \\ \hline
	\end{tabular}
	\caption{In the table presented above the Lowdin charges of all three systems are
		given. A change is observed in the Lowdin charges for complete and isolated system, this change due to both approximation in DFT and sharing of charges. The percentage difference gives a much clear picture, as the change for Benzene on Cu $\approx$ 10times higher than other two systems.
	}
	\label{Tab_lowdin_charges}
\end{table}

In conclusion, we have created artificial lattices based on fractal geometry  through adsorption of CO molecules on Cu(111) surface and studied via DFT.
Then LDOS and STM maps were computed for both artificial lattices.
Further, Hausdorff dimensions using box Minkowski-Bouligand method were calculated 
and observed to be fractions for both systems.
We also demonstrated that only Minkowski-Bouligand method is not sufficient for calculation of Hausdorff dimensions with Benzene on Cu and propose that Lowdin charge analysis should also be used.
Thus, demonstrating that fractals other than Sierpi\'nski triangle can be used to create 
artificial lattices.
We also demonstrated that one can study artificial fractal lattices and 
characterizing them using techniques such as DFT.

Additionally, the behaviour of fractional wavefunction of these lattices can be studied under the influence of an electrical field.
However, for the case of hexaflake no change in the fractal dimension is observed, see 
supplementary information for further details.
Such approach could open up several possibility of being utilized in quantum information sciences, 
where a system needs to be robust enough to be in entanglement while be probed.
Further, changes in wavefunction can be studied due to application of pressure on these systems and 
could be used for screening molecules like porphyrin on metal, as sensors.
Changes in the Hausdorff dimension due to application of stress-strain on lattices can also be 
studied.

\section{Acknowledgment}
The computational work described here is performed at the High Performance Computational 
Facility at IUAC, New Delhi, India.
We would like to express our gratitude to them.
Also, we would like to thank the University Grant Commission of India for providing partial 
funding for the research work through the UGC-BSR Research Startup Grant (Ref. 
No.F.30-309/2016(BSR)).

\end{document}